\documentclass[journal]{IEEEtran}
\usepackage{amsmath}
\usepackage{cases}

\usepackage{booktabs}

\usepackage{amssymb}
\usepackage{amsbsy}
\usepackage{amsmath}
\usepackage{bm}
\usepackage{verbatim}
\usepackage{cite}
\usepackage{mathrsfs}
\usepackage{amsfonts}
\usepackage{graphicx}
\usepackage[tight,footnotesize]{subfigure}
\usepackage[10pt]{moresize}
\usepackage{array}
\usepackage{color}
\usepackage{epsfig}
\usepackage{stfloats}
\usepackage{balance}

\usepackage{caption}

\usepackage[noend]{algpseudocode}

\usepackage{algorithmicx,algorithm}

\usepackage{cite}

\usepackage{setspace}

\usepackage{graphicx}
\usepackage{epstopdf}

\usepackage{xcolor}%¶¨ÒåÁËһЩÑÕÉ«
\usepackage{colortbl,booktabs}%µÚ¶þ¸ö°ü¶¨ÒåÁ˼¸¸ö*rule
\usepackage{tabularx}

\ifCLASSINFOpdf

\else

\fi

\begin{document}

\title{Channel-Correlation-Enabled Transmit Optimization for MISO Wiretap Channels}

\author{Shuai~Han,~\IEEEmembership{Senior~Member,~IEEE,}
        Sai~Xu,~\IEEEmembership{Student~Member,~IEEE,}\\
        Weixiao~Meng,~\IEEEmembership{Senior~Member,~IEEE,}
        and~Lei~He,~\IEEEmembership{Senior~Member,~IEEE}% <-this % stops a space
\thanks{Shuai~Han, Sai~Xu and Weixiao~Meng are with the Communications Research Center, Harbin Institute of Technology, China. (e-mail:
hanshuai@hit.edu.cn; fenicexusai@163.com; wxmeng@hit.edu.cn).}% <-this % stops a space
\thanks{Lei~He is with the Electrical Engineering Department, University of California,
Los Angeles, CA 90095, USA. (e-mail: lhe@ee.ucla.edu).}% <-this % stops a space
\thanks{Manuscript received XX XX, XXXX; revised XX XX, XXXX.}}

\maketitle

\begin{abstract}
An artificial-noise (AN)-aided beamformer specific to correlated main and wiretap channels is designed in this paper. We consider slow-fading multiple-input-single-output (MISO) wiretap channels with a passive single-antenna eavesdropper, in which independent transmitter-side and correlated receiver-side are assumed. Additionally, the source has accurate main channel information and statistical wiretap channel information. To reduce the secrecy loss due to receiver-side correlation, this paper proposes the scheme of channel-correlation-enabled transmit optimization. Particularly, the correlation is viewed as a resource to acquire more knowledge about wiretap channel. Based on this, the statistical distribution of wiretap channel is described more precisely. Then, the power of AN in the null space of main channel is placed more reasonably instead of simple uniform distribution and an elaborate beamformer for the information-bearing signal is designed. Finally, an efficient algorithm for power allocation between the information-bearing signal and the AN is developed. Simulation results show that the secrecy rate under transmit power and secrecy outage constraint is improved.
\end{abstract}


\IEEEpeerreviewmaketitle

\section{Introduction}

\IEEEPARstart{W}ith the Internet of Things (IoT) paradigm developing, an exponential increase of wireless devices along with many new applications beyond personal communications can be expected in the future \cite{S1-1,IoT-1}. A mass of access entities indisputably cause high device densities, which results in more serious security threats \cite{S1-2}. In addition to cryptography-based secrecy methods implemented at upper layers, physical layer security (PHY-security) techniques, by exploiting the random nature of physical layer transmission media to achieve both confidentiality and authentication \cite{S1-3}, also have attracted considerable attention. As an alternative to cryptography-based secrecy methods, PHY-security techniques, through proper coding and signal processing, ensure that confidential messages can be decoded with lower complexity only by the destination \cite{S1-4}. Compared to cryptography-based secrecy methods, PHY-security has some obvious advantages \cite{add-1}. For instance, PHY-security can guarantee information secrecy regardless of eavesdroppers' computational capability, which leads to that perfect secrecy can be achieved at the physical layer alone. Additionally, the centralized key distribution and management requested by cryptographic techniques can be eliminated by employing PHY-security techniques, with facilitating the management and improving the efficiency of wireless communications \cite{S1-5}.

\par In PHY-security, artificial-noise (AN)-aided beamforming techniques are typically utilized to strengthen secrecy. Beamforming aims to enhance the signal quality at the destination while limiting the signal strength at the eavesdropper; AN inserted into the transmit signal helps to degrade the reception at the eavesdropper. The authors in \cite{AN-1} concluded that transmitting the information-bearing signal in the direction of main channel is optimal to maximize the secrecy rate if only the statistics of wiretap channel are known. From the perspective of the average effect, this scheme is also applied to slow-fading channels. In \cite{AN-2}, AN was utilized to degrade the signal quality at the eavesdropper and was placed only in the null space of main channel with an isotropic distribution to avoid jamming the reception at the destination. The authors in \cite{AN-3} proved that AN in the null space of main channel must be symmetrical to maximize the secrecy when main and wiretap channels are independent of each other. When the desired signal was transmitted in the direction of main channel and AN was uniformly generated in the null space of main channel, the authors in \cite{AN-4} derived a closed-form expression for power allocation for minimizing the secrecy outage probability. Nevertheless, the scheme that full AN power is placed only in the null space of main channel is not optimal. This is because, when a part of AN power is placed in the direction of main channel, it is still possible that the net secrecy rate increases despite interference with the destination. The authors in \cite{AN-5} showed that the optimal AN is always orthogonal to the information-bearing signal for the maximum secrecy rate under the secrecy outage constraint and derived the optimal power allocation between the information-bearing signal and the AN without any approximation. In all these works, AN-aided beamforming techniques are discussed with the ideal assumption that main and wiretap channels are statistically independent of one another.

\par Some scenarios exist where main and wiretap channels are statistically correlated, that is, correlation occurs on receiver-side.  The correlation is harmful to the transmission security \cite{correlation-0}, and  excessively large signal power provides less benefit \cite{correlation-1}. In wireless environment, the extent of channel correlation often depends on some factors such as radio scattering, proximity and antenna deployments \cite{correlation-1}.
For example, antenna deployment at high altitudes in rural or suburban areas generates dominant line-of-sight paths, resulting in the correlation between the receive signals at two receivers. Moreover, it is also possible for the eavesdropper to actively induce the correlation by approaching the destination \cite{correlation-2}.
Considering that the aforementioned existing AN-aided beamforming schemes are not specific to the situation of correlated main and wiretap channels, the achievable secrecy may be not good enough. On the other hand, to reduce the secrecy loss due to correlation between main and wiretap channels, the channel fluctuation between antennas, users and relays is often exploited, which is followed by opportunistic selection implemented in a centralized or distributed manner through dedicated feedback links \cite{correlation-2,correlation-3}. For example, the authors in \cite{correlation-1} proposed that the secrecy could be enhanced by opportunistically transmitting messages in time slots instead of blindly increasing signal power. In particular, confidential transmission occurs when main channel has better instantaneous channel gain than that of wiretap channel. As another example, the authors in \cite{correlation-4} presented a scheme where transmit antenna selection is performed at the source, and the best relay is chosen for better transmission security. Further, the authors in \cite{correlation-5} studied the impact of both channel correlation and outdated relay selection on the secrecy performance. In addition to the strategy of opportunistic selection, a DCE scheme was proposed to enhance the secrecy for correlated main and wiretap channels in \cite{correlation-6}. In \cite{correlation-7}, AN wss emitted by the full-duplex destination in order to create interference to the eavesdropper. This strategy is highly desirable for correlated main and wiretap channels. In \cite{correlation-8}, multiple cooperative jammers were employed to interfere with the reception at the eavesdropper efficiently, aiming at reducing the disadvantage due to the correlation.

\subsection{Scope of Work}

% ´´Ðµã-------------------
% 1. ½¨Á¢¼ò½àÖ±¹ÛµÄMISOÏà¹ØÇÔÌýÐŵÀÄ£ÐÍ£¬²¢¸ø³öÎïÀíÒâÒåÉϵĽâÊÍ
% 2. ¶Ô²¨Êø³ÉÐÎʸÁ¿½øÐÐÁËÓÅ»¯Éæ¼°
% 3. ¶ÔÁã¿Õ¼äAN½øÐÐÁËÓÅ»¯Éæ¼°
% 4. ·¢Õ¹ÁËÒ»ÖÖ½ÏΪ¸ßЧµÄ¹¦ÂÊ·ÖÅäËã·¨
% ÓÅÊÆ----------------------
% ÒÔ±£ÃÜÖжϸÅÂÊϵÄÄ¿±ê±£ÃÜËÙÂÊΪĿ±êµÄAN¸¨Öú²¨Êø³ÉÐΣ¬
% ÏÖ´æµÄ×îºÃ·½°¸ÎªANÖÃÓÚÁã¿Õ¼ä£¬²¨Êø³ÉÐÎʸÁ¿Î»ÓÚÖ÷ÐŵÀ·½Ïò.
% ÎÒÃǵķ½°¸»ñµÃÁ˸üºÃµÄ±£ÃÜÐÔÄÜ

AN-aided beamforming design based on channel correlation for correlated multiple-input-single-output (MISO) wiretap channels has been partly studied in \cite{Myletter}, and the improvement and completion are conducted in this paper. Compared to the existing AN-aided beamforming strategies, the channel correlation information is utilized as a resource to help improve the secrecy in this paper. We seek to maximize the secrecy outage constrained secrecy rate by optimizing the AN-aided beamformer. The main contributions of our work are summarized as follows.

\begin{itemize}
\item[(1)] A simple and practical formula, describing the channel model of correlated MISO wiretap channels precisely, is derived. Through this formula, it is easy to observe that wiretap channel consists of both determinate and random components. Additionally, the relationship between main and wiretap channels is explained from the point of view of multipath effects. Based on this model, many problems in the scenario of correlated MISO wiretap channels can be conveniently studied.

\item[(2)] A channel-correlation-enabled AN-aided beamformer is designed to improve the secrecy for correlated MISO wiretap channels. Since the channel correlation suggests some useful information,  which can be utilized to estimate the statistical distribution of wiretap channel more precisely, the power of the AN in the null space of main channel is placed more reasonably instead of simple uniform distribution. Moreover, the beamformer for information-bearing signal can be optimized with better secrecy performance.

\item[(3)] An efficient algorithm for power allocation between the information-bearing signal and the AN is developed. Based on this, the achievable secrecy rate under the transmit power and secrecy outage constraint for correlated MISO channels can be obtained. By contrast, the secrecy performance the proposed suboptimal power allocation algorithm achieves approximates to that of the one-dimensional line brute-force search method with much less computing time.
\end{itemize}

\subsection{Paper Organization and Notations}

\par The remainder of this paper is organized as follows. In section II, the system model of correlated MISO wiretap channels is built. Based on this, the optimization problem is formulated to maximize the target secrecy rate under the secrecy outage probability and total power constraint. In section III, channel-correlation-enabled transmit optimization is studied, which involves correlation-based AN design, correlation-based beamforming design and power allocation between the information-bearing signal and the AN. Section IV presents simulations, and the numerical results demonstrate that the proposed channel-correlation-enabled transmit scheme can reduce the secrecy loss significantly for correlated MISO wiretap channels. Section V presents the conclusions of this work.

\par \emph{Notations}: Boldfaced lowercase and uppercase letters are used to represent vectors and matrices, respectively. All vectors are column vectors. The notations $( \cdot )^H$, $\otimes$, $\mathbb{E}[ \cdot ]$,
$\text{Tr}( \cdot )$, and $|\cdot|$ denote the conjugate transpose operation, the convolution operation, the mathematical expectation, the trace, the modulus of a complex number. $\left\|\cdot \right\|$ denotes the Euclidean norm of a vector or the Frobenius norm of a matrix. $\text{Re}(\cdot¡¤)$ is used to extract the real part of its argument.

\section{System Model}

\subsection{Channel Model}

\begin{figure}
\centering
\includegraphics[width=3.2in]{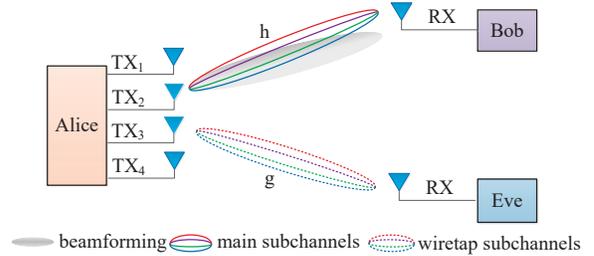}
\caption{Illustration of correlated main and wiretap channels.}
\label{Fig1}
\end{figure}

As illustrated in Fig. 1, we consider a multiple-input single-output single-antenna eavesdropper (MISOSE) system over slow-fading channels, where an $N_s$-antenna source (Alice) sends confidential messages to a single-antenna destination (Bob) with the transmission overheard by a passive single-antenna eavesdropper (Eve). Assuming that Alice has accurate channel state information (CSI) of the Alice-Bob link denoted by $\textbf{h}\in {{\textbf{C}}^{N_s \times 1}}$ and statistical CSI of the Alice-Eve link denoted by $\textbf{g}\in {{\textbf{C}}^{N_s \times 1}}$, with all elements of both channels following independent and identically distributed (i.i.d.) complex Gaussian distribution. Additionally, the transmitter-side is independent, whereas the receiver-side is correlated. In other words, the subchannels from the different transmit antennas to the same receive antenna are independent while those from the same transmit antenna to Bob and Eve are correlated. As shown in Fig. 1, the solid and dashed lines having the same colors characterize the correlation, while each color symbolizes the independence.

\par For the transmitter-side independence, it can be justified in the scenarios where the transmit antennas are surrounded by abundant scatterers and reflectors or the transmit antenna spacing is big enough to exploit the spatial multiplexing effect. For example, some micro base stations may be intentionally deployed in the position with abundant scatterers and reflectors in ultra-dense networks (UDNs). Mathematically, the transmitter-side independence is specified by $\textbf{h} \sim {\cal C}{\cal N}(0,{\sigma_d^2}{\textbf{I}_{N_s}})$ and $\textbf{g} \sim {\cal C}{\cal N}(0,{\sigma_e^2}{\textbf{I}_{N_s}})$ , where $\sigma_d^2$ and $\sigma_e^2$ are their channel gain variances, respectively. The receiver-side correlation occurs in the scenarios where Eve is around Bob and there are poor scatterers and reflectors locally around the two adjacent receivers. The correlation between channels from Alice to Bob and Eve can be characterized by the joint power probability density function (PDF) of their correlated SISO channels \cite{correlation-0}, that is,

\begin{equation}\label{eq1}
f_{h_i^2,g_i^2}(x,y) = \frac{{{I_0}(\frac{2}{{1 - \rho_i }}\sqrt {\frac{{\rho_i xy}}{{\sigma_d^2 \sigma _e^2}}} )}}{{(1 - \rho_i )\sigma_d^2 \sigma_e^2 }} {e^{ - \frac{\frac{x}{\sigma_d^2 } + \frac{y}{\sigma_e^2 }}
{1 - \rho_i }}},
\end{equation}

\noindent where $h_i$ and $g_i$ denote the subchannel gains from the \emph{i}-th antenna at Alice to Bob and Eve, respectively ($\emph{i} = 1,2,\cdots,N_s$). The function $I_0(x)$ is the modified Bessel function of the first kind of order zero \cite{correlation-9}, which is given by

\begin{equation}\label{eq2}
I_0(x) = \sum\limits_{k = 0}^{\infty} \left( \frac{x^k}{2^kk!} \right)^2.
\end{equation}

\noindent $\rho_i = \text{cov}(h_i^2,g_i^2)/\sqrt{\text{var}(h_i^2) \text{var}(g_i^2)}$ is the power correlation coefficient between main and wiretap subchannels. As Eve keeps passive near Bob, it is reasonably assumed that Eve is spatially randomly located around Bob. In a wireless environment, as a slow-varying statistical variable, the power correlation coefficient between main and wiretap subchannels in the specific regional scope around Bob can be estimated, with the maximum estimated value corresponding to $\rho_i$. Undoubtedly, $\rho_i \in [0,1)$ describes the degree of correlation in the worst case. In particular, $\rho_i$ = 0 denotes the independent subchannel scenario, while $|\rho_i|$ = 1 in the extreme case indicates that $h_{i}^2$ and $g_{i}^2$ are completely correlated. In conclusion, $\textbf{h}$ and $\textbf{g}$ consist of $N_s$ independent pairs of correlated SISO channels from Alice to Bob and Eve.

\par Since $h_i$ remains constant and $g_i$ as a variable is related to $h_i$ in a block, $g_{i}^2$ is conditional on $h_{i}^2$. Thus, the PDF of $g_{i}^2$ is determined by the joint power PDF of $h_{i}^2$ and $g_{i}^2$ divided by the PDF of $h_{i}^2$, that is,

\begin{equation}\label{eq3}
f_{g_i^2 | h_i^2} (x | y_i) = \frac{ f_{g_i^2,h_i^2} (x,y_i) }{ f_{h_i^2}(y_i) },
\end{equation}

\noindent where $f_{g_i^2 | h_i^2} (x | y_i)$ is the PDF of $g_{i}^2$ conditional on $h_{i}^2$ and $y_i$ is the instantaneous value of $h_i^2$ in this block. For $h_{i}^2$, it is a $\chi^2$ random variable with two degrees of freedom across a lot of blocks, whose PDF is

\begin{subnumcases}{f_{h_{i}^2}(y) =}
\frac{1}{\sigma_{d}^2} e^{-\frac{y}{\sigma_{d}^2}}, &$y > 0$ \\
\quad\quad 0\quad, &$y \leq 0$.
\end{subnumcases}

\noindent Thus,

\begin{subnumcases}{f_{g_i^2 | h_i^2} (x | y_i) =}
{\frac{{e^{ - \frac{{s_i^2 + x}}{{2\sigma _i^2}}}}}{{2\sigma _i^2}}{I_0}(\frac{{{s_i}}}{{\sigma _i^2}}\sqrt x )}, &$x > 0$ \\
\quad\quad\quad\quad 0\quad\quad\quad, &$x \leq 0$.
\end{subnumcases}

\noindent where

\begin{align}
\sigma _i^2 &= \frac{{(1 - \rho_i )\sigma_e^2}}{2},\\
{s_i} &= \sqrt {\frac{y_i \rho_i  \sigma_e^2}{\sigma_d^2}} .
\end{align}

\noindent From (5), it is not difficult to find that $g_i^2$ is a noncentral $\chi^2$ random variable with two degrees of freedom. Thus, the wiretap channel can be modeled as

\begin{equation}\label{eq8}
g_i = \frac{\sigma_e \sqrt {\rho_i}}{\sigma_d} h_i + \sigma_e \sqrt{1 - \rho_i } z,
\end{equation}

\noindent where $z$ is a zero-mean unit-variance complex Gaussian variable. Thus, it can be concluded that each element $g_i$ of wiretap channel is a complex Gaussian variable with its mean and variance being $\sigma_e \sqrt {\rho_i} /\sigma_d \cdot h_i $ and $\sigma_e^2 (1 - \rho_i)$ considering that $h_i$ is determinate in a block by the CSI feedback. From the point of view of multipath effects, the reason why $h_i$ and $g_i$ are correlated is that they have some common unsolvable paths. Let $\mathbf{\Theta} = \text{diag} \left\{\rho_1,\rho_2,\cdots,\rho_{N_s} \right\}$ and $\textbf{z} \sim {\cal C}{\cal N}(0,\textbf{I}_{N_s})$ . The wiretap channel is further formulated as

\begin{equation}\label{eq9}
\textbf{g} = \frac{\sigma_e}{\sigma_d} \widetilde{\textbf{h}} + \sigma_e \widetilde{\textbf{z}},
\end{equation}

\noindent where

\begin{align}
\widetilde{\textbf{h}} &= \sqrt{\mathbf{\Theta}} \textbf{h},\\
\widetilde{\textbf{z}} &= \sqrt{\mathbf{I}_{N_s}-\mathbf{\Theta}} \textbf{z}.
\end{align}

\subsection{Problem Formulation}

When Alice transmits a signal ${\textbf{x}}\in {{\textbf{C}}^{N_s \times 1}}$, the signals received at Bob and Eve are, respectively,

\begin{align}
y_d = \textbf{h}^H \textbf{x} + n_d,\\
y_e = \textbf{g}^H \textbf{x} + n_e,
\end{align}

\noindent where ${{n}_d}$ and ${{n}_e}$ denote the zero-mean unit-variance additive white complex Gaussian noises at Bob and Eve. To enhance the security, AN-aided beamforming scheme is employed for transmission. Since it is difficult to optimize the generalized AN design, we simplify the AN design by placing AN only in the null space of $\textbf{h}$. Specifically, the transmit signal vector $\textbf{x}$, consisting of both the message-bearing signal and the AN, is constructed as

\begin{equation}\label{eq11}
{\textbf{x}} = \sqrt {\phi P} \textbf{w} s  + \sqrt {(1-\phi) P} \textbf{a},
\end{equation}

\noindent where $\textbf{w}$ is a beamformer for transmitting the information-bearing signal \emph{s}. $P$ denotes the total transmit power and $\phi$ is the fraction of \emph{P} allocated to the information-bearing signal. \textbf{a} is the normalized AN vector in the null place of $\textbf{h}$ with $\textbf{h}^H\textbf{a} = 0$. Then, the SNRs at Bob and Eve are given by

\begin{align}
\delta_d &= \phi P \left| \textbf{h}^H \textbf{w} \right|^2,\\
\delta_e &= \frac{\phi P \left| \textbf{g}^H \textbf{w} \right|^2}{1 + (1-\phi) P \left| \textbf{g}^H \textbf{a} \right|^2}.
\end{align}

\par Given the received SNRs at Bob and Eve, the secrecy capacity over a block consisting of a large number of symbols is given by

\begin{subequations}\label {eq14}
\begin{align}
C_{s} = & \left[ C_{m}-C_{w} \right]^+,\\
=& \max \left\{ C_m - C_w,0 \right\},\\
=& \max \left\{ \log \left( \frac{1 + \delta_d}{1 + \delta_e} \right),0 \right\}.
\end{align}
\end{subequations}

\noindent In slow-fading channels, we use the secrecy outage probability to evaluate the secrecy performance of system. The secrecy outage probability is characterized as

\begin{equation}\label{eq15}
P_{out} \left(R_s \right) = \text{Pr} \left\{ C_s < R_s \right\},
\end{equation}

\noindent that is, the probability that the instantaneous secrecy capacity $C_s$ is less than the target secrecy rate $R_s$. Our objective is to maximize the target secrecy rate under the secrecy outage constraint. The problem is formulated as

\begin{subequations}\label {eq116}
\begin{align}
\underset{\textbf{a},\textbf{w},\phi} \max \quad\quad & R_s,\\
s.t. \quad\quad &p_{out} \left(R_s\right) \leq \varepsilon,\\
& 0 \leq \phi \leq 1,
\end{align}
\end{subequations}

\noindent where $\varepsilon \in [0,1)$ is a required secrecy outage probability.

\section{Channel-Correlation-Enabled Transmit Design}

\par Since the chance constrained AN-aided beamforming design (19) is non-convex, it is not likely to be solved efficiently. To make the problem tractable, we first find the suboptimal $\textbf{a}$ that determines the AN power distribution in the null space of $\textbf{h}$. Once $\textbf{a}$ is determined, the interference can be viewed as a constant for fixed $P$ and $\phi$. Then, the optimization of the beamformer $\textbf{w}$ for transmitting the information-bearing signal is studied. Finally,
we develop an efficient algorithm to find the suboptimal power allocation  between the
information-bearing signal and the AN. Through such three steps, the suboptimal solution to AN-aided beamforming design can be finalized. In this process, the channel correlation information is fully utilized as a resource to help improve the secrecy by describing the statistical distribution of wiretap channel more precisely.

\subsection{Correlation-Based AN Design}

The optimization problem (19) is non-convex and hard to be solved analytically. Even so, we can find the suboptimal scheme of AN power distribution in the null space of $\textbf{h}$. Specifically, for any fixed $P$ and $\phi$, according to Markov inequality \cite{Book},

\begin{equation}\label{eq17}
\text{Pr} \left\{ C_s \geq R_s \right\} \leq  \mathbb{E} \left[ C_s \right]/R_s.
\end{equation}

\noindent Thus,

\begin{equation}\label{eq18}
R_s \leq \frac{ \log\left(1+\delta_d\right) - \mathbb{E} \left\{ \log\left(1+\delta_e\right) \right\}}{1 - \varepsilon}.
\end{equation}

\noindent Clearly, the minimization of $\mathbb{E} \left\{ \log\left(1+\delta_e\right) \right\}$ contributes to the maximization of $R_s$.

\par Let

\begin{equation}\label{eq19}
\Phi(x) = \log\left[1 + \frac{\phi P {\left| \textbf{g}^H \textbf{w} \right|^2}}{1 + (1-\phi) P x } \right].
\end{equation}

\noindent The second derivative of the function $\Phi(x)$ is given by

\begin{align}
\Phi ''(x) =& \frac{\phi (1 - \phi )P^3 \left| \textbf{g}^H \textbf{w} \right|^2 } {\ln 2 \left[1 + (1 - \phi )Px \right]^2} \nonumber\\
&\times \frac{\phi (1 - \phi )P^3 \left| \textbf{g}^H \textbf{w} \right|^2 } {\left[ 1 + \phi P \left| \textbf{g}^H \textbf{w} \right|^2 + (1 - \phi )Px \right]^2 }.
\end{align}

\noindent Clearly, $\Phi ''(x) > 0$ must hold true for any $x$ with $\phi \in (0,1)$. Thus, $\Phi(x)$ is a convex function of $x$. According to Jensen's inequality, it is not difficult to find

\setcounter{equation}{20}
\begin{equation}\label {eq6}
\begin{aligned}
\Phi\left(\mathbb{E} [x]\right) \leq\mathbb{E}_x \left[\Phi(x)\right],
\end{aligned}
\end{equation}

\noindent where $\mathbb{E}_x [ \cdot ]$ denotes the mathematical expectation over $x$. Replacing $x$ with $\xi^2 = \left| \textbf{g}^H \textbf{a} \right|^2$, we observe that the increase of $ \mathbb{E} \left[\xi^2\right]$  helps reduce $\Phi\left(\mathbb{E} [\xi^2]\right)$, which is the lower bound of $\mathbb{E}_\xi \left[\Phi(\xi^2)\right] = \mathbb{E} \left[ \log\left(1+\delta_e\right) \right]$. As a result, the maximization of $R_s$ is facilitated.

\par Let $\rho_0 = \min \left\{\rho_1,\rho_2,\cdots,\rho_{N_s} \right\}$. According to (9),

\begin{equation}\label {eq6}
\textbf{g} = \frac{\sigma_e \sqrt{\rho_0}}{\sigma_d} \textbf{h} + \frac{\sigma_e }{\sigma_d} \widetilde{\textbf{h}} -  \frac{\sigma_e \sqrt{\rho_0}}{\sigma_d} \textbf{h} + \sigma_e \widetilde{\textbf{z}}.
\end{equation}

\noindent Considering $\textbf{h}^H\textbf{a} = 0$, it is easy to find

\begin{subequations}\label {eq9}
\begin{align}
\textbf{g}^H \textbf{a}
&=\widetilde{\textbf{g}}^H \textbf{a},\\
&= \left[ \frac{\sigma_e }{\sigma_d} \widetilde{\textbf{h}} -  \frac{\sigma_e \sqrt{\rho_0}}{\sigma_d} \textbf{h} + \sigma_e \widetilde{\textbf{z}} \right]^H \textbf{a}.
\end{align}
\end{subequations}

\noindent We observe that $\widetilde{g_i}$ denoting the the element of $\widetilde{\textbf{g}}$, is a complex Gaussian variable with its mean and variance being $\sigma_e \left(\sqrt {\rho_i}-\sqrt {\rho_0}\right)/\sigma_d \cdot h_i $ and $\sigma_e^2 (1 - \rho_i)$, Thus, the interference created by AN is given by

\begin{subequations}\label {eq14}
\begin{align}
\mathbb{E} \left\{\left| \textbf{g}^H \textbf{a} \right|^2\right\}
&=  \sum\limits_{i = 1}^{N_s} a_i^2 \mathbb{E} \left\{\widetilde{g_i}^2\right\},\\
&=  \sum\limits_{i = 1}^{N_s} a_i^2 (2 \sigma_i^2 + \widetilde{s_i}^2),
\end{align}
\end{subequations}

\noindent where $a_i$ is the element of \textbf{a} and $\widetilde{s_i}^2$ is given by

\begin{equation}\label {eq6}
\widetilde{s_i}^2 = \frac{ \sigma _e^2 (\sqrt{\rho_i} - \sqrt{\rho_0})^2h_i^2 }{\sigma_d^2}.
\end{equation}

\par Based on these discussions, we consider the following optimization problem.

\begin{subequations}\label {eq22}
\begin{align}
\underset{\textbf{a}} \max \quad\quad &J = \sum\limits_{i = 1}^{N_s} a_i^2 \mathbb{E} \left\{\widetilde{g_i}^2\right\},\\
s.t. \quad\quad &\textbf{h}^H \textbf{a} = 0,\\
&\text{Tr} (\textbf{a} \textbf{a}^H) = \sum\limits_{i = 1}^{N_s} a_i^2 \leq 1.
\end{align}
\end{subequations}

\noindent Although the optimization problem (26) is non-convex, we can apply augmented Lagrangian method to find its solution. For (26), the augmented Lagrangian is constructed as

%its local extreme point as

\begin{align}
L (\textbf{a},\mu,\nu) &= J + \frac{1}{2\kappa } \left\{ \left[ \max \left\{ 0, \mu + \kappa \text{Tr} (\textbf{a} \textbf{a}^H)   \right\} \right])^2 -  \mu^2 \right\} \nonumber \\
& \quad\quad\quad\quad\quad\quad\quad \quad\quad + \nu \textbf{h}^H \textbf{a} + \frac{\kappa }{2} |\textbf{h}^H \textbf{a}|^2,
\end{align}

\noindent where $\mu$ and $\nu$ are the multipliers and $\kappa$ is the penalty parameter. The formulas used to update the multipliers are given by

\begin{align}
\nu_{k+1} &= \nu_{k} + \kappa \textbf{h}^H \textbf{a}_{k},\\
\mu_{k+1} &= \max \left\{ 0, \mu_{k} + \kappa \text{Tr} (\textbf{a}_{k} \textbf{a}_{k}^H) \right\},
\end{align}

\noindent and the judgement function is given by

\begin{equation}\label{eq19}
\phi_k (\textbf{a}) = \left\{ | \textbf{h}^H \textbf{a}_{k}|^2 + \left[ \max \left\{ -\text{Tr} (\textbf{a}_{k} \textbf{a}_{k}^H), \frac{\mu_{k}}{\kappa} \right\} \right]^2 \right\}^{1/2}.
\end{equation}

\noindent The procedure to solve the problem (26) is summarized in Algorithm 1.

\begin{algorithm}[t]
\caption{Optimization of AN}
\hspace*{0.02in} {\bf Input:}
$\rho_i, \sigma_d, \sigma_e, \textbf{h}$, etc.;\\
\hspace*{0.02in} {\bf Output:} %Ëã·¨µÄ½á¹ûÊä³ö
$\textbf{a}$;
\begin{algorithmic}[1]
\State Initialize: $\textbf{a}, \mu, \nu, \kappa, \varepsilon, T, t = 1, 0< \theta <1, c>1$;
\Repeat
\State  $p = \phi (\textbf{a})$;
\State $\textbf{a} \leftarrow \arg \underset{\textbf{a}} \min \left\{ L (\textbf{a},\mu,\nu) \right\}$;
\If{$\phi (\textbf{a}) < \epsilon$}
\State break;
\EndIf
\If{$\phi (\textbf{a}) \geq \theta \times p $}
\State $\kappa = \min \left\{1000, c\times\kappa \right\};$
\EndIf
\State $\nu\leftarrow \nu + \kappa \textbf{h}^H \textbf{a}$;
\State $\mu \leftarrow \max \left\{ 0, \mu + \kappa \text{Tr} (\textbf{a} \textbf{a}^H) \right\};$
\State $t \leftarrow t+1$;
\Until{$t< T$}
\end{algorithmic}
\end{algorithm}

\par By a series of processing in this subsection, the suboptimal AN vector \textbf{a} in the null space of $\textbf{h}$ is obtained. It is not difficult to find that \textbf{a} only depends on \textbf{h} and the statistical distribution of \textbf{g}, and has nothing with the beamformer $\textbf{w}$ for transmitting the information-bearing signal. Clearly, this result makes it easy to design the beamformer $\textbf{w}$.

\subsection{Correlation-Based Beamforming Design}

For fixed $P$ and $\phi$, with AN power distribution in the null space of $\textbf{h}$ is obtained, the chance constrained optimization problem (19) is simplified into

\begin{subequations}\label {eq11}
\begin{align}
\underset{\textbf{w}} \max \quad\quad & R_s,\\
s.t. \quad\quad &p_{out} \left(R_s\right) \leq \varepsilon,\\
& \textbf{w}^H\textbf{w}\leq1.
\end{align}
\end{subequations}

\noindent However, the problem (31) is still non-convex and is not likely to be solved directly. According to \cite{{w-1},{w-2}}, the problem (31) is equivalently reformulated as a probability constrained power minimization problem with the target rate $R_s > 0$ as follows.

\begin{subequations}\label {eq11}
\begin{align}
\underset{\textbf{w}} \min \quad\quad & \textbf{w}^H\textbf{w},\\
s.t. \quad\quad &p_{out} \left(R_s\right) \leq \varepsilon.
\end{align}
\end{subequations}

\noindent For any given target secrecy rate $R_s$, the optimal solution $\textbf{w}^{*}$ must exist. Besides, the optimal objective value $\textbf{w}^{*H}\textbf{w}^{*} $ of the above problem is monotonically increasing with respect to $R_s$. Taking the two points into account, the target secrecy rate $R_s^*$ enabling the optimal objective value $\textbf{w}^{*H}\textbf{w}^{*} = 1$ can be found by using a bisection search over different $R_s$. Clearly, $R_s^*$ must be the optimal solution to the problem (31). The procedure of searching $R_s^*$ is summarized in Algorithm 2.

\begin{algorithm}[t]
\caption{Bisection Search}
\hspace*{0.02in} {\bf Input:}
$\varepsilon$, $\phi$, \emph{P}, \textbf{h}, \textbf{a}, $f_{{g_i}} (x)$, etc.;\\
\hspace*{0.02in} {\bf Output:} %Ëã·¨µÄ½á¹ûÊä³ö
$R_s^*$;
\begin{algorithmic}[1]
\State Initialize: $\epsilon$, $R_u$ and $R_l$;
\Repeat
\State $R_s^*  \leftarrow (R_u + R_l)/2$;
\If{$p_{out} \left(R_s^* \right) \leq \varepsilon$}
\If{$\textbf{w}^H\textbf{w}\leq1$}
\State $R_l = R_s^* $;
\Else
\State $R_u = R_s^* $;
\EndIf
\Else
\State $R_u = R_s^* $;
\EndIf
\Until{$|R_u - R_l| < \epsilon $}
\end{algorithmic}
\end{algorithm}

\par In the following, we focus on the probability constrained power minimization problem with the given target rate $R_s$. Similar to \cite{{w-1}}, we employ the approach of relaxation-restriction to solve this problem: the chance constraint is conservatively transformed into a deterministic form (the restriction step) and then an semidefinite relaxation (SDR) is performed to lift the problem into a high dimension (the relaxation step).

\subsubsection{Conservative Transformation} For the chance constraint (32b), it can be reformulated as

\begin{align}
p_{out} \left(R_s\right) &= \Pr\Bigg\{ \log \left(1 + \phi P \left| \textbf{h}^H \textbf{w} \right|^2 \right)  \nonumber \\
& - \log \left[ 1 + \frac{\phi P \left| \textbf{g}^H \textbf{w} \right|^2}{1 + (1-\phi) P \left| \textbf{g}^H \textbf{a} \right|^2} \right] \le {R_s} \Bigg\}.
\end{align}

\noindent Further, it can be rewritten as

\begin{equation}\label{eq13}
\Pr \left\{ \omega  < \phi P \left| \textbf{g}^H \textbf{w} \right|^2 - (1-\phi)P\omega \left| \textbf{g}^H \textbf{a} \right|^2\right\} ,
\end{equation}

\noindent where

\begin{align}
& \omega  = 2^{-R_s} \left( 1 + \phi P \left| \textbf{h}^H \textbf{w} \right|^2 \right)  - 1,\\
& \left| \textbf{g}^H \textbf{a} \right|^2 = \sum\limits_{i = 1}^{N_s} a_i^2 \widetilde{g_i}^2.
\end{align}

\noindent From (9), we observe that the wiretap channel $\textbf{g}$ consists of two parts: the deterministic component ${\sigma_e}/{\sigma_d} \cdot\widetilde{\textbf{h}}$ and the stochastic component $\sigma_e \cdot \widetilde{\textbf{z}}$. Substituting (9) into $\left| \textbf{g}^H \textbf{w} \right|^2$, it can be rewritten as

\begin{subequations}\label {eq11}
\begin{align}
\left| \textbf{g}^H \textbf{w} \right|^2 &= \left(\frac{\sigma_e}{\sigma_d} \widetilde{\textbf{h}} + \sigma_e \widetilde{\textbf{z}} \right)^H \textbf{w} \textbf{w}^H \left(\frac{\sigma_e}{\sigma_d} \widetilde{\textbf{h}} + \sigma_e \widetilde{\textbf{z}}  \right), \\
&= \frac{\sigma_e^2 }{\sigma_d^2} \widetilde{\textbf{h}}^H \textbf{w}\textbf{w}^H \widetilde{\textbf{h}} + \sigma_e^2 \widetilde{\textbf{z}}^H \textbf{w}\textbf{w}^H \widetilde{\textbf{z}} \nonumber \\
&\quad\quad\quad\quad\quad\quad\quad\quad + \frac{2\sigma_e^2 }{\sigma_d} \text{Re} \left( \widetilde{\textbf{z}}^H \textbf{w}\textbf{w}^H \widetilde{\textbf{h}}\right).
\end{align}
\end{subequations}

\noindent  Similarly, $\left| \textbf{g}^H \textbf{a} \right|^2$ can be rewritten as

\begin{align}\label{eq47}
\left| \textbf{g}^H \textbf{a} \right|^2
&= \frac{\sigma_e^2 }{\sigma_d^2} \widetilde{\textbf{h}}^H \textbf{a}\textbf{a}^H \widetilde{\textbf{h}} + \sigma_e^2 \widetilde{\textbf{z}}^H \textbf{a}\textbf{a}^H \widetilde{\textbf{z}} \nonumber \\
&\quad\quad\quad\quad\quad\quad\quad\quad + \frac{2\sigma_e^2 }{\sigma_d} \text{Re} \left( \widetilde{\textbf{z}}^H \textbf{a}\textbf{a}^H \widetilde{\textbf{h}}\right).
\end{align}

\noindent Noting that $\textbf{a} \textbf{a}^H $ is a diagonal matrix because $\textbf{a}$ is a random  Gaussian noise vector.
Thus, (34) can be rewritten as

\begin{equation}\label{eq13}
\Pr \Bigg\{ \omega - \frac{\sigma_e^2 }{\sigma_d^2} \widetilde{\textbf{h}}^H \textbf{A} \widetilde{\textbf{h}}  <   \textbf{z}^H \mathbf{\Lambda} \textbf{z} + 2 \text{Re} \left( \textbf{z}^H \textbf{x}\right) \Bigg\} ,
\end{equation}

\noindent where

\begin{align}
 &\mathbf{A}= \phi P \textbf{w}\textbf{w}^H - (1-\phi)P\omega \textbf{a}\textbf{a}^H,\\
 &\mathbf{\Lambda}= {\sigma_e^2 } \sqrt{\mathbf{I}_{N_s}-\mathbf{\Theta}} \mathbf{A} \sqrt{\mathbf{I}_{N_s}-\mathbf{\Theta}} ,\\
 &\textbf{x}= \frac{ \sigma_e^2 }{\sigma_d}\sqrt{\mathbf{I}_{N_s}-\mathbf{\Theta}} \mathbf{A} \widetilde{\textbf{h}}.
\end{align}

\newtheorem{lemma}{Lemma}
\begin{lemma} [Bernstein-Type Inequality I] \label{lemma1}
Let $\textbf{G}=\textbf{z}^H \mathbf{\Lambda} \textbf{z} + 2 \text{Re} \left( \textbf{z}^H \textbf{x}\right) $ , where $\mathbf{\Lambda}\in {{\textbf{C}}^{N \times N}}$ is a complex Hermitian matrix, $\textbf{z} \sim \mathcal{CN}(0, \textbf{I}_N)$ and $\textbf{x} \in {{\textbf{C}}^{N \times 1}}$. Then for any $\sigma \geq 0$, we have

\begin{align}
\Pr  \Bigg\{ \textbf{G} \geq \text{Tr} \left( \mathbf{\Lambda}\right)+ \sqrt{2 \sigma} \sqrt{ \parallel \text{vec}\left( \mathbf{\Lambda}\right)\parallel^2 + 2 \parallel \textbf{x} \parallel^2}  \nonumber\\
+ \sigma s^+ \left( \mathbf{\Lambda}\right)  \Bigg\} \leq \text{exp}(-\sigma),
\end{align}

% vec±íʾÏòÁ¿»¯£¬½«¾ØÕóÀ­Ö±¡£https://en.wikipedia.org/wiki/Vectorization_(mathematics)
% ÀýÈ磺A = [a b; c d]Ôòvec(A) = [a c b d]^H

\noindent where $s^+ \left( \mathbf{\Lambda}\right)= \max \left\{\lambda_{\max}\left( \mathbf{\Lambda}\right),0 \right\} $ with $\lambda_{\max}\left( \mathbf{\Lambda}\right)$ denoting the maximum eigenvalue of matrix $\mathbf{\Lambda}$.

\end{lemma}

\par Clearly, \emph{Lemma 1} can bound the tail probability of quadratic forms of Gaussian variables involving matrices. According to \emph{Lemma 1} and (39), the chance constraint (32b) can be conservatively transformed into the following deterministic form:

\begin{align}
\text{Tr} \left( \mathbf{\Lambda}\right)+ \sqrt{2 \sigma} \sqrt{ \parallel \text{vec}\left( \mathbf{\Lambda}\right)\parallel^2 + 2 \parallel \textbf{x} \parallel}
+ \sigma s^ + \left( \mathbf{\Lambda}\right) \nonumber\\
\leq  \omega - \frac{\sigma_e^2 }{\sigma_d^2} \widetilde{\textbf{h}}^H \textbf{A} \widetilde{\textbf{h}},
\end{align}

\noindent where $\sigma=-\ln(\varepsilon)$. In other words,  if (44) is true, then the chance constraint the chance constraint (32b) must hold true. Consequently, the relaxed problem (32) is now conservatively reformulated as

\begin{subequations}\label {eq11}
\begin{align}
\underset{\textbf{w}} \min \quad\quad & \textbf{w}^H\textbf{w},\\
s.t. \quad\quad &(44).
\end{align}
\end{subequations}

\subsubsection{Semidefinite Relaxation}

Define $\textbf{W} \overset{\vartriangle}{=} \textbf{w}\textbf{w}^H$. It is easy to find that (45) is equivalent to

\begin{subequations}\label {eq11}
\begin{align}
\underset{\textbf{w}} \min \quad\quad & \text{Tr}\left(\textbf{W}\right),\\
s.t. \quad\quad &(44),\\
&\textbf{W}\succeq 0, \text{rank}(\textbf{W})=1.
\end{align}
\end{subequations}

\noindent The problem (46) can be relaxed again by using the semidefinite relaxation (SDR) approach to drop the rank constraint $\text{rank}(\textbf{W})=1$. The rank relaxed problem becomes

\begin{subequations}\label {eq11}
\begin{align}
\underset{\textbf{w}} \min \quad\quad & \text{Tr}\left(\textbf{W}\right),\\
s.t. \quad\quad &(44), \textbf{W} \succeq 0.
\end{align}
\end{subequations}

\noindent It is easy to see that the above problem is equivalent to the following problem

\begin{subequations}\label {eq11}
\begin{align}
\underset{\textbf{w}} \min \quad\quad & \text{Tr}\left(\textbf{W}\right),\\
s.t. \quad\quad &\text{Tr} \left( \mathbf{\Lambda}\right)+ \sqrt{2 \sigma} \alpha + \sigma \beta - c \leq 0,\\
& c = \omega - \frac{\sigma_e^2 }{\sigma_d^2} \widetilde{\textbf{h}}^H \textbf{A} \widetilde{\textbf{h}},\\
& {\left\| {\begin{array}{*{20}{c}}
{{\rm{vec}}\left( {\mathbf{\Lambda }} \right)}\\
{\sqrt 2 {\textbf{x}}}
\end{array}} \right\| \le \alpha {\rm{ }}},\\
& \beta\text{I}-\mathbf{\Lambda}\succeq 0, \beta >0, \textbf{W} \succeq 0,
\end{align}
\end{subequations}

\noindent where $\alpha$ and $\beta$ are slack variables. This problem has a linear objective, a second order cone constraints and two convex PSD constraints. Therefore, it is a convex problem and can be solved by using off-the-shelf convex optimization solvers. However, due to the rank relaxation, there is no guarantee that the resulting optimal solution $\textbf{W}$ is feasible for the original problem (46). To obtain a feasible rank-one solution $\widehat{\textbf{w}}$, we employ a simple Projection Approximation Procedure which is proposed in \cite{w-1}. Specifically, the feasible $\widehat{\textbf{w}}$ can be given by the following three steps:

\begin{itemize}
\item[i.]  Let \textbf{P} denote the project matrix of vector $\textbf{W}^{1/2}\widetilde{\textbf{h}}$, that is,
\begin{equation}\label{eq13}
\textbf{P} = \dfrac{\textbf{W}^{1/2}\widetilde{\textbf{h}} \left(\textbf{W}^{1/2}\widetilde{\textbf{h}}\right)^H}{ \parallel \textbf{W}^{1/2} \widetilde{\textbf{h}} \parallel^2};
\end{equation}
\item[ii.]  We construct a new rank-one solution
\begin{equation}\label{eq13}
\widehat{\textbf{W}} = \textbf{W}^{1/2} \textbf{P} \textbf{W}^{1/2};
\end{equation}
\item[iii.]
Since $\widehat{\textbf{W}}$ is a rank-one complex Hermitian matrix, we can obtain $\widehat{\textbf{w}}$ from $\widehat{\textbf{W}}$ By SVD method. Specifically,

\begin{subequations}\label {eq9}
\begin{align}
\widehat{\textbf{W}} =& \textbf{Q} \mathbf{\Sigma} \textbf{Q}^{H}\\
=& (\textbf{Q} \mathbf{\Sigma}^{1/2} ) \left(\textbf{Q}\mathbf{\Sigma}^{1/2}\right)^{H}\\
=& \widehat{\textbf{w}}\widehat{\textbf{w}}^H.
\end{align}
\end{subequations}
\end{itemize}

\noindent According to \cite{w-1}, this simple scheme is guaranteed to find a rank-one solution which has performance no worse than $\textbf{W}^\ast$.

\subsection{Power Allocation}

%\subsection{Suboptimal Power Allocation between Signal and AN}

\par For any fixed $P$ and $\phi$, the AN power distribution in the null space of $\textbf{h}$ and the correlation-based beamformer can be determined by the previous two subsections. In the following, we focus on optimizing power allocation between the information-bearing signal and the AN. Although the optimal power allocation coefficient can be obtained by a one-dimensional line brute-force search within [0, 1], it is inefficient. Thus, we will develop an efficient algorithm.

\par According to (21), we observe that $R_s$ can be enlarged by maximizing

\begin{equation}\label{eq13}
C(\phi) = \log\left(1+\delta_d\right) - \mathbb{E} \left\{ \log\left(1+\delta_e\right) \right\}.
\end{equation}

\noindent Clealy, the rate paid for preventing being eavesdropped is $\mathbb{E} \left\{ \log\left(1+\delta_e\right) \right\}$. Let

\begin{align}\label{eq47}
H(\phi) = &\log\left(1+\phi P \left| \textbf{h}^H \textbf{w} \right|^2 \right)    \nonumber \\
& \quad -\log\left\{1+ \frac{\phi P  \left| \textbf{g}^H \textbf{w} \right|^2 }{1 + (1-\phi) P \left| \textbf{g}^H \textbf{a} \right|^2 } \right\}.
\end{align}

\noindent Clearly, $H(\phi)$ can be viewed as a reference guide for the variation trend of $C(\phi)$. For $H(\phi)$, its first derivative is given by

\begin{equation}\label{eq48}
H'(\phi) = H'_1(\phi) - H'_2(\phi),
\end{equation}

\noindent where

\begin{align}
&H'_1(\phi)= \frac{1}{\ln 2} \frac{P \left| \textbf{h}^H \textbf{w} \right|^2}{1 + \phi P \left| \textbf{h}^H \textbf{w} \right|^2},\\
&H'_2(\phi)= \frac{1}{\ln 2} \frac{P \left| \textbf{g}^H \textbf{w} \right|^2  \left(1 + P \left| \textbf{g}^H \textbf{a} \right|^2  \right)}{ 1 + (1- \phi) P   \left| \textbf{g}^H \textbf{a} \right|^2 } \nonumber\\
&\times \frac{1}{1 + (1- \phi) P  \left| \textbf{g}^H \textbf{a} \right|^2  + \phi P  \left| \textbf{g}^H \textbf{w} \right|^2  }.
\end{align}

\par From (53), we observe that the first and second terms of $H(\phi)$ both rise with the increase of $\phi$. According to (55) and (56),  the denominators of their derivatives $H'_1(\phi)$ and $H'_2(\phi)$ are the linear function and the quadratic function of $\phi$, respectively. In the situation of (53), there may exist two cases: i) $H'_1(\phi)$ declines continually with $\phi$ increasing while $H'_2(\phi)$
rises first then declines with $\phi$ increasing (When $H'_2(\phi)$ declines, its rate is below $H'_1(\phi)$);  ii) both of $H'_1(\phi)$ and $H'_2(\phi)$ decline with $\phi$ increasing and the decline rate of $H'_1(\phi)$ is higher than $H'_2(\phi)$. In either case, it can be concluded that $H'(\phi)$ is monotonically decreasing with respect to $\phi$ and there exists only one extreme point for $H(\phi)$ within $\phi\in (0,1)$. These results cursorily reflect that $\phi$ corresponding to the maximum value of $C(\phi)=0$ is confined to a small range. Therefore, for the problem (19), the power allocation coefficient $\phi^\ast$ between the information-bearing signal and the AN can be obtained by the following three steps: i) the root $\phi_0$ of $C'(\phi)=0$ is extracted by bisection method. Clearly, $\phi_0$ is the value of $\phi$ at the rough extreme point of $C(\phi)$. ii) $\phi_{0}$ can be used as an initial value to execute brute-force search within a narrow range around $\phi_{0}$ so as to find a better power allocation coefficient $\phi^\ast$ for the problem (19).

\par To compute the achievable secrecy rate, $f_{\delta_e}(x)$ representing the PDF of $\delta_e$ is necessary. Let $\zeta = \textbf{g}^H \textbf{w} $. $\zeta$ can be expanded into

\begin{subequations}\label {eq11}
\begin{align}
\zeta &=  \sum\limits_{i = 1}^{N_s} \left( g_{ix} - j g_{iy}\right) \left( w_{ix} + j w_{iy} \right),\\
&=\sum\limits_{i = 1}^{N_s}  g_{ix} w_{ix} + g_{iy} w_{iy} + j \sum\limits_{i = 1}^{N_s} g_{ix} w_{iy} - g_{iy} w_{ix}.
\end{align}
\end{subequations}

\noindent Considering that $g_i$ is a complex Gaussian variable with its mean and variance being $\sigma_e \sqrt {\rho_i} /\sigma_d \cdot h_i $ and $\sigma_e^2 (1 - \rho_i)$, $\sum\nolimits_{i = 1}^{N_s}  g_{ix} w_{ix} + g_{iy} w_{iy}$ is also a complex Gaussian variable with its mean and variance being $\sigma_e /\sigma_d \cdot \text{Re} \left( \widetilde{\textbf{h}}^H\textbf{w} \right)=\sigma_e /\sigma_d \cdot \sum\nolimits_{i = 1}^{N_s} \sqrt {\rho_i} \left( h_{ix} w_{ix} + h_{iy} w_{iy}\right)$ and $\sigma_e^2/2 \cdot |\sqrt{\mathbf{I}_{N_s}-\mathbf{\Theta}}\textbf{w}|^2 =\sigma_e^2/2 \cdot \sum\nolimits_{i = 1}^{N_s} (1-\rho_i) \left(w_{ix}^2 + w_{iy}^2\right)$. Similarly, $\sum\nolimits_{i = 1}^{N_s} g_{ix} w_{iy} - g_{iy} w_{ix}$ is a complex Gaussian variable with its mean and variance being $\sigma_e /\sigma_d \cdot\text{Im} \left( \widetilde{\textbf{h}}^H\textbf{w} \right)=\sigma_e /\sigma_d \cdot \sum\nolimits_{i = 1}^{N_s} \sqrt {\rho_i} \left( h_{ix} w_{ix} - h_{iy} w_{iy}\right)$ and $\sigma_e^2/2 \cdot |\sqrt{\mathbf{I}_{N_s}-\mathbf{\Theta}}\textbf{w}|^2 =\sigma_e^2/2 \cdot \sum\nolimits_{i = 1}^{N_s} (1-\rho_i) \left(w_{ix}^2 + w_{iy}^2\right)$. Thus, $\zeta^2= \left| \textbf{g}^H \textbf{w} \right|^2$ is a noncentral $\chi^2$ random variable with two degrees of freedom. Thus, the PDF of $\zeta^2$ is given by

\begin{subnumcases}{f_{\zeta^2} (x) =}
{\frac{{e^{ - \frac{{s_\zeta^2 + x}}{{2\sigma _\zeta^2}}}}}{{2\sigma _\zeta^2}}{I_0}(\frac{{{s_\zeta}}}{{\sigma _\zeta^2}}\sqrt x )}, &$x > 0$ \\
\quad\quad\quad\quad 0\quad\quad\quad, &$x \leq 0$.
\end{subnumcases}

\noindent where

\begin{align}
\sigma_\zeta^2 &= \frac{\sigma_e^2}{2} \cdot |\sqrt{\mathbf{I}_{N_s}-\mathbf{\Theta}}\textbf{w}|^2,\\
s_\zeta &= \frac{\sigma_e }{\sigma_d} \cdot \left| \widetilde{\textbf{h}}^H\textbf{w} \right|.
\end{align}

\noindent According to (36), $\xi^2 = \left| \textbf{g}^H \textbf{a} \right|^2 = \sum\nolimits_{i = 1}^{N_s} a_i^2 \widetilde{g_i}^2$, while $\widetilde{g_i}$ and $\widetilde{g_j}$ ($i\neq j$) are independent of one another. Thus, we can compute the PDF of $\xi^2$ by the characteristic function method. Specifically, according to the previous conclusion that $\widetilde{g_i}$ is a complex Gaussian variable with its mean and variance being $\sigma_e \left(\sqrt {\rho_i}-\sqrt {\rho_0}\right)/\sigma_d \cdot h_i $ and $\sigma_e^2 (1 - \rho_i)$, the characteristic function of $F_{ \widetilde{g_i}^2} (x)$ denoting the CDF of $\widetilde{g_i}^2$ is given by

\begin{equation}\label{eq38}
\Phi_{\widetilde{g_i}^2}(\varpi) = \left( \frac{1}{1-2 j \varpi \sigma_i^2} \right) e^{\frac{j \varpi \widetilde{s_i}^2}{1-2 j \varpi \sigma_i^2}}.
\end{equation}

\noindent Thus, the characteristic function of $F_{\xi^2} (x)$ denoting the CDF of $\xi^2$ is given by

\begin{equation}\label{eq38}
\Phi_{\xi^2}( \varpi) =  \prod_{i=1}^{N_s} \Phi_{\widetilde{g_i}^2}(a_i^2 \varpi).
\end{equation}

\noindent Then, the PDF of $\xi^2$ is given by

\begin{equation}\label {eq11}
f_{\xi^2}(x) =  \lim_{T \rightarrow \infty}  \frac{1}{2 \pi} \int_{-T}^T e^{-j \varpi x} \Phi_{\xi^2}(\varpi) d\varpi.
\end{equation}

\par Let $\mu = \phi P {\zeta^2}$ and $\nu = 1 + (1-\phi) P \xi^2 $. The PDFs of $\mu$ and $\nu$ are given by

\begin{align}
f_{\mu}(x) &= \frac{f_{\zeta^2} (x)(\frac{x}{\phi P})}{\phi P},\\
f_{\nu}(x) &=  \frac{f_{\xi^2 }\left(\frac{x}{(1-\phi) P}-\frac{1}{(1-\phi) P}\right)}{(1-\phi) P}.
\end{align}

%\begin{algorithm}[t]
%\caption{Correlation-Based Beamforming Algorithm} %�㷨������
%\hspace*{0.02in} {\bf Input:} %�㷨�����룬 \hspace*{0.02in}��������λ�ã�ͬʱ���� \\ ���л���
%$\varepsilon$, \emph{P}, \textbf{h}, $\mathbb{E} \left[g_i \right]$, etc.;\\
%\hspace*{0.02in} {\bf Output:} %�㷨�Ľ�������
%$R_s^\ast$;
%\begin{algorithmic}[1]
%\State Initialize: search step size $\Delta=0.001$, required accuracy $\kappa = 0.001$, $\phi=0$, $R_s=0$, optimal $R_s^\ast=0$;% \State ��дһ������
%\State Solve (14) and obtain \textbf{a};
%\Repeat
%\State Set initial bounds $R_u=\log (1+ P \left| \textbf{h} \right|^2 )$ and $R_l=0$;
%\Repeat
%\State Update $R_s \leftarrow (R_u + R_l)/2$;
%\If{$F_{\delta_e}\left({(\delta_d + 1)/2^{R_s}-1}\right) > 1-\varepsilon$} % If ���䣬��Ҫ��EndIf��Ӧ
%����\State $R_l = R_s$;
%\Else
%����\State $R_u = R_s$;
%\EndIf
%\Until{$|R_u - R_l| < \kappa$}
%\If{$R_s>R_s^\ast$}
%\State $R_s^\ast \leftarrow R_s$;
%\EndIf
%\State $\phi \leftarrow \phi + \Delta$;
%\Until{$\phi > 1$}
%\State \Return $R_s^\ast$.
%\end{algorithmic}
%\end{algorithm}

\noindent On this basis, the CDF of $\delta_e = \frac{\mu}{\nu}$ can be computed as follows.

\begin{subequations}\label{eq13}
\begin{align}
F_{\delta_e}(x)&= \text{Pr} \left( \delta_e \leq x \right)\\
&= \text{Pr} \left( \frac{\mu}{\nu} \leq x \right)\\
&= \text{Pr} \left( \mu \leq \nu x, \nu>1 \right) \\
&= \int_1^\infty  {\left( {\int_{0}^{\nu x} {{f_\mu}(\mu)d\mu} } \right)} {f_\nu}(\nu)d\nu,
\end{align}
\end{subequations}

%\noindent Then, $f_{\delta_e}(x)$ can be given by
%
%\begin{equation}\label{eq38}
%f_{\delta_e}(x) = \frac{d F_{\delta_e}(x)}{dx}
%\end{equation}

\noindent From Problem (19), we observe that the maximum $R_s$ must occur in the extreme case of $p_{out} \left(R_s\right) = \varepsilon$. Thus,

\begin{equation}\label{eq38}
\varepsilon = 1-F_{\delta_e}({\frac{\delta_d + 1}{2^{R_s}}-1}).
\end{equation}

%\begin{align}
%\varepsilon &= 1-\int_{ 0 }^{\frac{\delta_d + 1}{2^{R_s}}-1} f_{\delta_e}(x) dx\\
%&= 1-F_{\delta_e}({\frac{\delta_d + 1}{2^{R_s}}-1}).
%\end{align}

\noindent By solving (67), the achievable secrecy rate $R_s$ for any $\phi$ within a narrow range around $\phi_{0}$ can be obtained. The maximum value $R_s^\ast$ among all these $R_s$ is the suboptimal solution to the problem (19). Now the procedure of our proposed AN-aided beamforming design based on correlation is summarized in Algorithm 3.

\par Note that although the closed-form solution to $f_{\xi^2}(x)$ and $F_{\delta_e}(x)$ are not available, we can compute it quickly via numerical integration. Up to present, the proposed channel-correlation-enabled transmit design has been introduced thoroughly.

\begin{algorithm}[t]
\caption{Secrecy Rate under Secrecy Outage Constraint}
\hspace*{0.02in} {\bf Input:}
$\varepsilon$, \emph{P}, \textbf{h}, $\mathbb{E} \left[g_i \right]$, etc.;\\
\hspace*{0.02in} {\bf Output:}
$R_s^\ast$;
\begin{algorithmic}[1]
\State Initialize: $R_s^\ast=0$, $\Delta$, $\epsilon$;
\State Solve (26) to obtain \textbf{a};
\State Solve (46) to obtain \textbf{W};
\State Find $\widehat{\textbf{w}}$ to (32) by Projection Approximation Procedure;
\State Find $R_s^\ast$ and the corresponding \textbf{w} by Algorithm 1;
\State Solve (53) to obtain $\phi_{0}$ by bisection method;
\State $\phi \leftarrow \phi_{0}-\Delta/2$;
\Repeat
\State Solve (67) to obtain $R_s$ by bisection method under $\epsilon$;
\State $R_s^\ast \leftarrow \max\left\{R_s^\ast,R_s\right\}$;
\State $\phi \leftarrow \phi +\epsilon$;
\Until $\phi > \phi_{0}+\Delta/2$
\end{algorithmic}
\end{algorithm}

% https://wenku.baidu.com/view/16059072cdbff121dd36a32d7375a417866fc1a8.html

%˼·��
%1�����ʷ���ϵ�������󲻳����������ⲿ��һ����������һ�¼��ɣ������ܶȿ���ʹ�����ٷ��������Ż��㷨��
%2�������ܶ��ȸ���һ�����������������ܾ����ˣ���һ��ʹ�����Ϲ���Ҳ���������պϸ����ܶȽ⣬������������ֵ�⡣
%3����Ϊû�и����ܶȱպϽ⣨�����п���Ҳ���ˣ���Rs����������ʹ�ö��ַ������˵������ӡ�
%4�������������ܶȿ��Կ���ʹ�ö���ʽ���ϱ���ʽ��

%�����ܶȿ��Ը��ݾ��������������ٶ��ұ���ʽ.
%�����������ն�ż��KKT����
%���߶��ַ�������������������
%�򻯷��������ռ�ANʹ������ֵ������ͨ���󵼱����ҳ����ѵ��źź�AN���ʷ���

\section{Numerical Results}

In this section, computer simulations are performed to evaluate the secrecy performance achieved by the proposed channel-correlation-enabled AN-aided beamforming scheme.

\subsection{Validating Channel Model}

\begin{table}
\begin{center}
\caption{Simulation parameters for validating the correctness of the correlated wiretap channel model.}~~\\
\setlength{\tabcolsep}{3mm}{
\begin{tabular}{cccccc} \toprule
       & $\beta_{d}$  &  $\beta_{e}$  &  $\rho$  &$h^2$ &$S$  \\ \hline
case 1  & 1  & 1 & 0.5  & 0.1535 & $10^6$ \\
case 2  & 1  & 1 & 0.3  & 0.2826& $10^6$ \\
case 3  & 1  & 1 & 0.7  & 1.6469  & $10^6$ \\
case 4  & 0.5  & 0.5 & 0.5  & 0.4681 & $10^6$ \\
\bottomrule
\end{tabular}}
\end{center}
\end{table}

\begin{figure}
\centering
\includegraphics[width=3.2in]{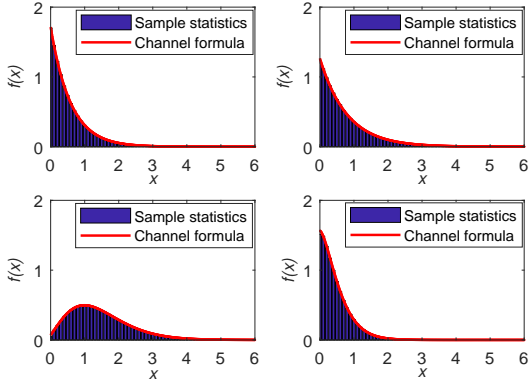}
\caption{Validating the correctness of the correlated wiretap channel model.}
\label{Fig1}
\end{figure}

Before illustrating the secrecy performance, we first validate the correctness of the channel model succinctly formulating the relationship between main and wiretap channels by simulated data. In the simulation, four cases are given, which involved the parameters the channel gain variances $\sigma_d^2$ and $\sigma_e^2$, the power correlation coefficient $\rho$ between main and wiretap channels, and the sample size $S$. For the main channel gain in a time slot, it can be generated by sampling from Rayleigh distribution. The power of main channel gain and the parameters are listed in Table I. Fig. 2 presents the probability distribution of wiretap channel according to (8) and the statistical result of simulated data which are sampled by (1). It is clear that the statistical result roughly agrees with the probability distribution, which validates the correctness of the proposed channel model.

\subsection{Secrecy Performance}

\begin{table}
\begin{center}
\caption{Simulation parameters.}~~\\
\setlength{\tabcolsep}{3mm}{
\begin{tabular}{cccccccc} \toprule
       & $\sigma_{d}^2$  &  $\sigma_{e}^2$  &  $\rho$     &$\varepsilon$ &$P$ &$N_s$ \\ \hline
$R_s$ \emph{vs.} $\rho$  & 1  & 1 & -    &0.15 &10dBW &8\\
$R_s$ \emph{vs.} $P$   & 1  & 1 & 0.5    &0.15 &- &8\\
$R_s$ \emph{vs.} $\varepsilon$  & 1  & 1 & 0.5     &- &10dBW &8\\
$R_s$ \emph{vs.} $N_s$  & 1  & 1 & 0.5    &0.15 &10dBW &-\\
\bottomrule
\end{tabular}}
\end{center}
\end{table}

We start by investigate how the correlation coefficient between main and wiretap channels affects the secrecy rate under the transmit power and secrecy outage constraint. Then, the secrecy rate under the secrecy outage constraint with respect to the total transmit power is given. Next, the secrecy rate under the transmit power constraint versus the secrecy outage probability are presented. Finally, how the secrecy rate under the transmit power and secrecy outage constraint depends on the number of the transmit antennas is shown. In simulations, the solid curve represents the proposed channel-correlation-enabled AN-aided beamforming scheme, while the red dashed curve represents the traditional scheme that the information-bearing signal is transmitted in the direction of main channel and AN is uniformly embedded into the null space of main channel. It is worth mentioning that, for independent main and wiretap channels, when the optimization objective is the secrecy rate under the transmit power and secrecy outage constraint, the traditional scheme is the best existing one. Meanwhile, the brute-force scheme corresponding to the blue dashed curve is also presented. Compared to the proposed scheme, the only difference is that the brute-force scheme employs brute-force search to obtain the power allocation coefficient.

\par In simulations, the correlation coefficient between main and wiretap channels is measured by $\rho$, which indicates the expectation of channel correlation in wireless environment. And the correlation for each subchannel from Alice to Bob and Eve is assumed to uniformly fluctuate between $[\rho - 0.2, \rho +0.2]$.

\subsubsection{$R_s$ \emph{vs.} $\rho$}

\begin{figure}
\centering
\includegraphics[width=3.2in]{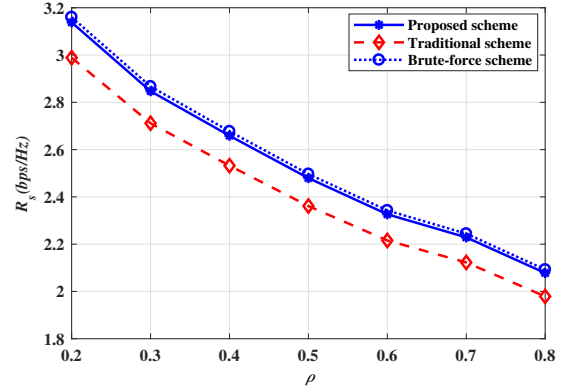}
\caption{The relationship between the power correlation coefficient $\rho$ and the achievable secrecy rate $R_s$.}
\label{Fig1}
\end{figure}

Fig. 3 shows the relationship between the power correlation coefficient $\rho$ and the achievable secrecy rate $R_s$. As shown in TABLE II, some parameters in the simulation are set as follows: the number of antennas at Alice is $N_s = 8$; the required secrecy outage probability is $\varepsilon$ = 0.15; the variances of $\textbf{h}$ and $\textbf{g}$, related to the path loss as a function of distance, are set as $\sigma_{d}^2 = \sigma_{e}^2 = 1$; and the total transmit power $P$ = 10 dBW. We observe that $R_s$ depends largely on $\rho$. As $\rho$ increase, $R_s$ declined rapidly. That means that the correlation between main and wiretap channels is harmful to the transmission security, especially when the correlation is high. This result is consistent with our experience and knowledge. We also observe that the proposed scheme achieves greater percent gains of $R_s$ with $\rho$ increasing compared to the traditional scheme. The reason is that, higher correlation contributes to acquiring more knowledge about wiretap channel and the transmit optimization specific to correlated main and wiretap channels becomes definitely more effective. On the other hand, the achievable performance of the proposed scheme is very close to the brute-force scheme, which verifies the usefulness of the proposed algorithm of power allocation, which works almost as well as brute-force search with much lower computational complexity.

\subsubsection{$R_s$ \emph{vs.} $P$}

\begin{figure}
\centering
\includegraphics[width=3.2in]{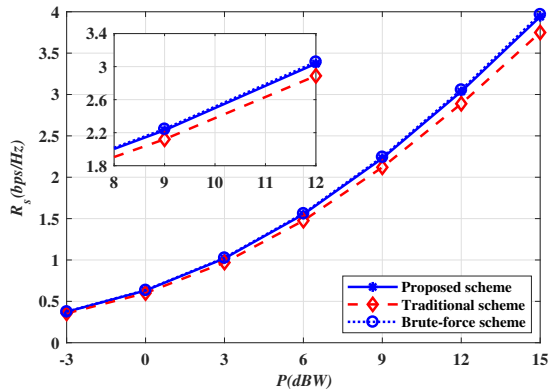}
\caption{The relationship between the total transmit power $P$ and the achievable secrecy rate $R_s$.}
\label{Fig1}
\end{figure}

Fig. 4 shows the relationship between the total transmit power $P$ and the achievable secrecy rate $R_s$. As shown in TABLE II, some parameters in the simulation are set as follows: the number of antennas at Alice is $N_s = 8$; the required secrecy outage probability is $\varepsilon$ = 0.15; the variances of $\textbf{h}$ and $\textbf{g}$, related to the path loss as a function of distance, are set as $\sigma_{d}^2 = \sigma_{e}^2 = 1$; and the power correlation coefficient $\rho$ = 0.5. It can be clearly seen that $R_s$ depends largely on $P$. With $P$ increasing, $R_s$ becomes large. Despite such a fact, the excessively large power provides less benefit. For example, when $P$ changes from 12 dBW to 15 dBW, the power increases by half but $R_s$ only grows by approximately 25\%. On the other hand, the proposed scheme achieves higher $R_s$ with $\rho$ increasing compared to the traditional scheme. It indicates that the channel correlation can be regarded as a resource to the optimize the transmission so as to further strengthen security. Like the previous simulation, the achievable performance of the proposed scheme is very close to the brute-force scheme with $P$ increasing, which reconfirms that the proposed algorithm of power allocation works almost as well as brute-force search.

\subsubsection{$R_s$ \emph{vs.} $\varepsilon$}

\begin{figure}
\centering
\includegraphics[width=3.2in]{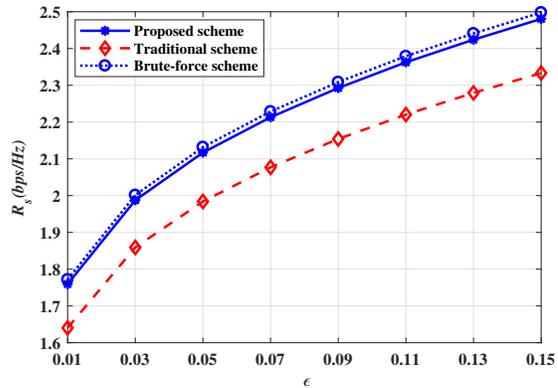}
\caption{The relationship between the secrecy outage probability $\varepsilon$ and the achievable secrecy rate $R_s$.}
\label{Fig1}
\end{figure}

Fig. 5 shows the relationship between the secrecy outage probability $\varepsilon$ and the achievable secrecy rate $R_s$. As shown in TABLE II, some parameters in the simulation are set as follows: the number of antennas at Alice is $N_s = 8$; the variances of $\textbf{h}$ and $\textbf{g}$, related to the path loss as a function of distance, are set as $\sigma_{d}^2 = \sigma_{e}^2 = 1$; and the power correlation coefficient $\rho$ = 0.5; and  and the total transmit power $P$ = 10 dBW. We observe that an increase in $\varepsilon$ causes a rapid increase in $R_s$, which conforms to practical secrecy communications. Moreover, the proposed scheme grows faster than the traditional scheme, which means that optimizing the  beamformer for the information-bearing signal and the power distribution of AN in the null space of main channel can facilitate the secrecy. In accordance with the previous two simulation results, the achievable performance of the proposed scheme is very close to the brute-force scheme and works almost as well as brute-force search with much lower computational complexity.

\subsubsection{$R_s$ \emph{vs.} $N_s$}

\begin{figure}
\centering
\includegraphics[width=3.2in]{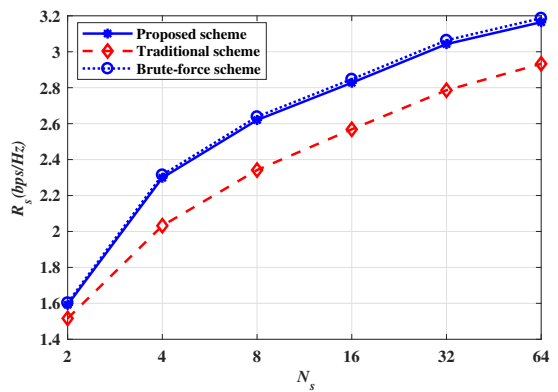}
\caption{The relationship between the number of transmit antenna $N_s$ and the achievable secrecy rate $R_s$.}
\label{Fig1}
\end{figure}

Fig. 6 shows the relationship between the number of transmit antenna $N_s$ and the achievable secrecy rate $R_s$. As shown in TABLE II, some parameters in the simulation are set as follows: the required secrecy outage probability is $\varepsilon$ = 0.15; the variances of $\textbf{h}$ and $\textbf{g}$, related to the path loss as a function of distance, are set as $\sigma_{d}^2 = \sigma_{e}^2 = 1$; and the power correlation coefficient $\rho$ = 0.5; and the total transmit power $P$ = 10 dBW. We observe that an increase in $N_s$ contributes to improving $R_s$. The reason is that the increase of antennas can provide more spatial degrees of freedom. By comparison, it is not difficult that, the secrecy rate of the proposed scheme is always above that of the traditional scheme. In accordance with the previous three simulation results, the achievable performance of the proposed scheme is very close to the brute-force scheme and works almost as well as brute-force search with much lower computational complexity. These simulation results confirm the superiority, effectiveness and feasibility of the proposed scheme together.

\section{Conclusions}

This paper studies channel-correlation-enabled AN-aided beamforming design for correlated MISO wiretap channels, aiming at maximizing the secrecy rate under the transmit power and secrecy outage constraint. By exploiting the correlation, the wiretap channel is formulated as the sum of both determinate and random components, which describes wiretap channels precisely and succinctly. Based on this, the power of AN in the null space of main channel is placed more reasonably instead of simple uniform distribution. And the beamformer for information-bearing signal is optimized specific to the scenario of correlation. Then, an efficient algorithm for power allocation between the information-bearing signal and the AN is developed. Simulation results show that the secrecy rate under the transmit power and secrecy outage constraint is improved significantly.

%Note that the trade-off for this performance gain may be in the form of longer delays, higher system complexity, restrictive scenarios, etc.

% or


\ifCLASSOPTIONcaptionsoff
  \newpage
\fi


\begin{thebibliography}{1}
\bibitem{S1-1}
M. Mohammadi, A. Al-Fuqaha, S. Sorour, and M. Guizani , ``Deep Learning for IoT Big Data and Streaming Analytics: A Survey,'' \emph{IEEE Communications Surveys \& Tutorials}, vol. 20, no. 4, pp. 2923-2960, 2018.
\bibitem{IoT-1}
M. Agiwal, A. Roy, and N. Saxena , ``Next Generation 5G Wireless Networks: A Comprehensive Survey,'' \emph{IEEE Communications Surveys \& Tutorials}, vol. 18, no. 3, pp. 1617-1655, 2016.
\bibitem{S1-2}
S. Han, S. Xu, W. Meng, and C. Li, ``Dense-Device-Enabled Cooperative Networks for Efficient and Secure Transmission,'' \emph{IEEE Network}, vol. 32, no. 2, pp. 100-106, 2018.
\bibitem{S1-3}
Y. Liu, H. Chen, and L. Wang, ``Physical Layer Security for Next Generation Wireless Networks: Theories, Technologies, and Challenges,'' \emph{IEEE Communications Surveys \& Tutorials}, vol. 19, no. 1, pp. 347-376, 2017.
\bibitem{S1-4}
S. Han, S. Xu, W. Meng, and C. Li, ``An Agile Confidential Transmission Strategy Combining Big Data Driven Cluster and OBF,'' \emph{IEEE Transactions on Vehicular Technology}, vol. 66, no. 11, pp. 10259-10270, 2017.
\bibitem{add-1}
N. Yang, L. Wang, G. Geraci, M. Elkashlan, J. Yuan, and M. Di Renzo,``Safeguarding 5G wireless communication networks using physical layer security,'' \emph{IEEE Communications Magazine}, vol. 53, no. 4, pp. 20-27, Apr. 2015
\bibitem{S1-5}
S. Yan, X. Zhou, N. Yang, B. He and T. D. Abhayapala,``Artificial-Noise-Aided Secure Transmission in Wiretap Channels With Transmitter-Side Correlation,'' \emph{IEEE Transactions on wireless Communications}, vol. 15, no. 12, Dec. 2016.


\bibitem{AN-1}
P. Lin, S. Lai, S. Lin, and H. Su, ``On Secrecy Rate of the Generalized Artificial-Noise Assisted Secure Beamforming for Wiretap Channels,'' \emph{IEEE Journal on Selected Areas in Communications}, vol. 31, no. 9, pp. 1728-1740, Sep. 2013.
\bibitem{AN-2}
S. Goel, and R. Negi, ``Guaranteeing Secrecy Using Artificial Noise,'' \emph{IEEE Transactions on wireless Communications}, vol. 7, no. 6, pp. 2180-2189, 2008.
\bibitem{AN-3}
S. Gerbracht, C. Scheunert, and E. Jorswieck, ``Secrecy Outage in MISO Systems with Partial Channel Information,'' \emph{IEEE Transactions on Information Forensics and Security}, vol. 7, no. 2, pp. 704-716, Apr. 2012.
\bibitem{AN-4}
J. Xiong, K. Wong, D. Ma, and J. Wei, ``A Closed-Form Power Allocation for Minimizing Secrecy Outage Probability for MISO Wiretap Channels via Masked Beamforming,'' \emph{IEEE Communications Letters}, vol. 16, no. 9, pp. 1496-1499, Sep. 2012.

\bibitem{AN-5}
Wang, Bo, P. Mu, and Z. Li, ``Secrecy Rate Maximization with Artificial-Noise-Aided Beamforming for MISO Wiretap Channels under Secrecy Outage Constraint,'' \emph{IEEE Communications Letters}, vol. 19, no. 1, pp. 18-21, 2015.

\bibitem{correlation-0}
X. Sun, J. Wang, W. Xu, and C. Zhao, ``Performance of Secure Communications over Correlated Fading Channels,'' \emph{IEEE Signal Processing Letter}, vol. 19, no. 8, pp. 479-482, Aug. 2012.
\bibitem{correlation-1}
H. Jeon, N. Kim. J. Choi, H. Lee, and J. Ha, ``Bounds on Secrecy Capacity over Correlated Ergodic Fading Channels at High SNR,'' \emph{IEEE Transactions on Information Theory}, vol. 75, no. 4, pp. 1975-1983,
\bibitem{correlation-2}
L. Fan, X. Lei, T. Q. Duong, M. Elkashlan, and G. K. Karagiannidis, ``Secure Multiuser Communications in Multiple Amplify-and-Forward Relay Networks,'' \emph{IEEE Transactions on Communications}, vol. 62, no. 9, pp. 3299-3310, Sep. 2014.
\bibitem{correlation-3}
N. Yang, P. L. Yeoh, M. Elkashlan, R. Schober, and J. Yuan, ``MIMO Wiretap Channels: Secure Transmission Using Transmit Antenna Selection and Receive Generalized Selection Combining,'' \emph{IEEE Communications Letter}, vol. 17, no. 9, pp. 1754-1757, Sep. 2013.
\bibitem{correlation-4}
L. Fan, R. Zhao, F. Gong, N. Yang, and G. K. Karagiannidis,`` Secure Multiple Amplify-and-Forward Relaying over Correlated Fading Channels,'' \emph{IEEE Transactions on Communications}, vol. 65, no. 7, pp. 2811- 2820, 2017.
\bibitem{correlation-5}
L. Fan, X. Lei, N. Yang, T. Q. Duong, and G. K. Karagiannidis, ``Secrecy Cooperative Networks with Outdated Relay Selection over Correlated Fading Channels.'' \emph{IEEE Transactions on Vehicular Technology}, vol. 66, no. 8, pp. 7599-7603, Aug. 2017.
\bibitem{correlation-6}
Y. Du, S. Han, S. Xu, and C. Li. ``Improving Secrecy under High Correlation via Discriminatory Channel Estimation,'' in \emph{IEEE International Conference on Communications}, Kansas City, MO, USA, May 2018, pp. 1-6.
\bibitem{correlation-7}
G. Zheng, I. Krikidis, J. Li, A. P. Petropulu, and B. Ottersten, ``Improving Physical Layer Secrecy Using Full-Duplex Jamming Receivers,'' \emph{IEEE Transactions on Signal Processing}, vol. 61, no. 20, pp. 4962-4974, Oct. 2013.
\bibitem{correlation-8}
S. Xu, S. Han, W. Meng, Y. Du, and L. He, ``Multiple-Jammer-Aided Secure Transmission With Receiver-Side Correlation,'' \emph{IEEE Transactions on Wireless Communications}, vol. 18, no. 6, pp. 3093-3103, 2019.

\bibitem{Myletter}
S. Xu, S. Han, Y . Du, W. Meng, L. He, and C. Zhang, ``AN-Aided Secure Beamforming Design for Correlated MISO Wiretap Channels,'' \emph{IEEE Communications Letters}, vol. 23, no.4, pp. 628-631, 2019.

\bibitem{correlation-9}
I. S. Gradshteyn and I. M. Ryzhik, \emph{Table of Integrals, Series, and Products}, 7th ed. San Diego, CA, USA: Academic, 2007

\bibitem{Book}
J. G. Proakis and M. Salehi, \emph{Digital Communications}, 5th ed. New York: McGraw-Hill, 2008.

\bibitem{w-1}
S. Ma, M. Hong, E. Song, X. Wang, and D. Sun, ``Outage Constrained Robust Secure Transmission for MISO Wiretap Channels,'' \emph{IEEE Transactions on Wireless Communications}, vol. 13, no. 10, pp. 5558-5570, Oct. 2013.
\bibitem{w-2}
J. Xu, S. Xu, and C. Xu, ``Outage Constrained Robust Secure Transmission for MISO Wiretap Channels,'' \emph{IEEE Access}, vol. 5, pp. 10277-10284, 2017.






%\bibitem{other-2}
%L. Fan, X. Lei, et al., `` Secrecy Cooperative Networks with Outdated Relay Selection over Correlated Fading Channels,'' \emph{IEEE Trans. Veh. Technol.}, vol. 66, no. 8, pp. 7599-7603, Aug. 2017.

\end{thebibliography}
\end{document}